\documentclass[10pt,showpacs,twocolumn,amsmath,amssymb,english,aps,twocolumn]{revtex4}
\usepackage[T1]{fontenc}
\usepackage[latin9]{inputenc}
\usepackage{amsthm}
\usepackage{amsmath}
\usepackage{dcolumn}
\usepackage{amssymb}
\usepackage{graphicx}
\usepackage{esint}
\usepackage{epsfig}

\makeatletter

\@ifundefined{textcolor}{}
{%
 \definecolor{BLACK}{gray}{0}
 \definecolor{WHITE}{gray}{1}
 \definecolor{RED}{rgb}{1,0,0}
 \definecolor{GREEN}{rgb}{0,1,0}
 \definecolor{BLUE}{rgb}{0,0,1}
 \definecolor{CYAN}{cmyk}{1,0,0,0}
 \definecolor{MAGENTA}{cmyk}{0,1,0,0}
 \definecolor{YELLOW}{cmyk}{0,0,1,0}
 }
\numberwithin{equation}{section}
\numberwithin{figure}{section}
  \theoremstyle{plain}
  \newtheorem*{lem*}{\protect\lemmaname}
 \theoremstyle{definition}
 \newtheorem*{defn*}{\protect\definitionname}
  \theoremstyle{definition}
  \newtheorem*{example*}{\protect\examplename}
  \theoremstyle{plain}
  \newtheorem*{prop*}{\protect\propositionname}
  \theoremstyle{plain}
  \newtheorem*{conjecture*}{\protect\conjecturename}
  \theoremstyle{plain}
  \newtheorem*{thm*}{\protect\theoremname}

\makeatother

\usepackage{babel}
  \providecommand{\conjecturename}{Conjecture}
  \providecommand{\definitionname}{Definition}
  \providecommand{\examplename}{Example}
  \providecommand{\lemmaname}{Lemma}
  \providecommand{\propositionname}{Proposition}
  \providecommand{\theoremname}{Theorem}

\begin{document}

\title{Approximate entropy of network parameters}

\author{James West}

\email{jawest@gmail.com}

\selectlanguage{english}%

\affiliation{Statistical Cancer Genomics, Paul O\textquoteright{}Gorman Building,
UCL Cancer Institute}
\affiliation{Department of Physics \& Astronomy, University College London}

\author{Lucas Lacasa}
\email{lucas.lacasa@upm.es }
\selectlanguage{english}%
\affiliation{Departamento de Matemática Aplicada y Estadística, ETSI Aeronáuticos, Universidad Politécnica de Madrid}

\author{Simone Severini}
\email{simoseve@gmail.com}
\selectlanguage{english}%
\affiliation{Department of Computer Science}
\affiliation{Department of Physics \& Astronomy, University College London }

\author{Andrew Teschendorff}
\email{a.teschendorff@ucl.ac.uk }
\selectlanguage{english}%
\affiliation{Statistical Cancer Genomics, Paul O\textquoteright{}Gorman Building,
UCL Cancer Institute }

\begin{abstract}
We study the notion of approximate entropy within the framework of network theory. Approximate entropy is an uncertainty measure originally proposed in the context of dynamical systems and time series. We firstly define a purely structural entropy obtained by computing the approximate entropy of the so called \emph{slide sequence}. This is a surrogate of the degree sequence and it is suggested by the frequency partition of a graph. We examine this quantity for standard scale-free and Erd\H{o}s\textendash{}Rényi networks. By using classical results of Pincus, we show that our entropy measure converges with network size to a certain binary Shannon entropy. On a second step, with specific attention to networks generated by dynamical processes, we investigate approximate entropy of horizontal visibility graphs. Visibility graphs permit to naturally associate to a network the notion of temporal correlations, therefore providing the measure a dynamical garment. We show that approximate entropy distinguishes visibility graphs generated by processes with different complexity. The result probes to a greater extent these networks for the study of dynamical systems. Applications to certain biological data arising in cancer genomics are finally considered in the light of both approaches.
\end{abstract}
\pacs{89.75.Hc, 89.75.Fb, 89.75.-k, 89.70.Cf}
\maketitle

\section{Introduction}
Concepts such as information, entropies, and measures of complexity are highly connected subjects within the core of dynamical systems and chaos theory.
The area benefits from a wealth of literature that dates back to the works of Kolmogorov on metric entropies, and continues with the developments
of Sinai, Eckmann, Ruelle, and others (see \cite{Eck85} and references therein for a review on the topic). Roughly speaking, this branch of science is relatively mature for answering questions such as how system, which is sensitive to initial conditions (with positive characteristic Lyapunov exponents), generates uncertainty as time evolves, and how this entropy production is related to the structure (invariant measure) of the system.

In recent years, in parallel with the advent of the study of complex networks (see, for example, \cite{New03,New10}), similar ideas aiming to describe the amount of organization of these systems are starting to take root. As a matter of fact, to describe mathematically the amount of heterogeneity and complexity found in natural and technological networks is nowadays a major endeavor in the frameworks of network theory, general data analysis, and inference.
Several recent works point towards an entropic origin for a variety of key properties of complex networks that we find around us, such as the biodiversity maintenance in ecological networks \cite{Bast09, Basc06}, or, more generally, the emergence of robust degree-degree correlations \cite{Joh10} and communities in social and biological networks \cite{Bia09bis}. Indeed, the amount of heterogeneity in a network is a basic ingredient for quantifying properties of diffusion processes, like the spread of human epidemics, computer viruses, \emph{etc.} \cite{Bog03,Sat01,Sat04}.

Some theoretical approaches to deal with the notion of network heterogeneity include the references \cite{Bia09,Bia08}, where a statistical mechanics perspective is adopted to estimate the (thermodynamic) entropy of network ensembles given by a set of constraints. Other lines of research make use of spectral theory to derive optimal network configurations \cite{Don05, BGS, Pass08}. However, the majority of proposed network-based entropic functionals are, so far, entropies \emph{\`{a} la} Shannon (see also \cite{de11} for a recent review). Fewer work has been reported on the extension of other invariant measures to the network theoretic context.

In order to contribute filling this gap, in this paper we consider the notion of \emph{approximate entropy} as introduced by Pincus \cite{Pin91}. This is a finite size statistic of the Eckmann-Ruelle entropy originally proposed as a measure to determine the complexity of a system that changes in time. Time series are in fact the main area of application.
We explore network-based extensions of the approximate entropy. We ask questions about the usefulness of this parameter to estimate degrees of uncertainty both in the case of static and growing networks. It is important to see that any attempt of defining an approximate entropy of network parameters (following the same line of Pincus) must be based on some ordering of the data. As a starting point, it is natural to consider certain orderings associated with the degree sequence and its related quantities.

The present work contains several contributions: (1) the general point is the novelty of studying the notion of approximate entropy in a network theoretic context; (2) this requires the translation of network parameters into a time series that is amenable to be investigated with analytic tools, accordingly, we define a binary string associated to the degree sequence as it is suggested by the notion of frequency partition; (3) as expected, given that the string reflects some coarse grained properties of the degree sequence, we are able to show that the approximate entropy of the string distinguishes between common network ensembles; (4) we then consider visibility graphs, since these objects are associated to dynamical systems and therefore present a natural time ordering. We show that approximate entropy of those visibility graphs allows to distinguish between series generated by different types of processes.\\

The remainder of the paper is organized as follows. In Section II we recall the definition and main properties of approximate entropy. In Section III we study how to extend such notion to the context of networks, defining a network-based approximate entropy. Given this measure, we study both the case of static networks (Section IV), where the measure is purely structural, and the case of growing networks (Section V), where the measure acquires a more dynamical meaning. This latter situation is studied within the context of visibility graphs \cite{Lac08, Luq09}. The measure is finally tested in Section VI with real data. By considering networks constructed from data obtained in the context of cancer statistics, we probe the capability of the measure to distinguish amongst different cancer phenotypes. A discussion is finally presented in Section VII. The main open question concerns the study of approximate entropies of parameters beyond the degrees. One may take a number of different approaches depending on the parameters considered. Sequences of combinatorial nature obtained from counting paths and sequences of algebraic nature, like for example, graph spectra, seem to be the good candidates for this purpose. Concerning the latter idea, it is an open direction to determine whether approximate entropy has any role in characterizing matrix ensembles when applied to their spectra.

\section{Approximate entropy}
We begin by recalling the notion of approximate entropy. Due to Pincus \cite{Pin91}, its definition is based on ideas of Eckmann, Ruelle, and ultimately Kolmogorov-Sinai. Whereas originally defined for time series, when the series is drawn from an alphabet of finitely many symbols,
it has a powerful combinatorial interpretation due to Rukhin \cite{Ruk00}. However, approximate entropy has in general a geometric interpretation
given when comparing {}``densities'' of the Takens embedding of the time
series in dimensions $m$ and $m+1$. Indeed, let $m\in\mathbb{N}$,
$r\in\mathbb{R}_{>0}$, $u=\left(u(t)\right)_{t=1}^{N}$ a time series
of $N$ points and consider the $m$th Takens embedding delay map
$x(t)=\left(u(t),u(t+1),...,u(t+m-1)\right)$,
with image $X_{m}=\left\{ x(1),...,x\left(n\right)\right\} \subset\mathbb{R}^{m}$,
where $n=N-\left(m-1\right)$. Recall that if $u$ arises from a dynamical
system with a strange attractor of box dimension $d$ then, when $m>2d$,
the image $X_{m}$ {}``reconstructs'' the strange attractor in an
appropriate sense. We write $x_{i}(t)$ for the $i$th component of
$x\in\mathbb{R}^{m}$ and let $\left\Vert \cdot\right\Vert _{\infty}$
be the usual $L^{\infty}$ norm on $\mathbb{R}^{m}$, \emph{i.e.} $\left\Vert x\right\Vert _{\infty}=\max_{i}|x_{i}|$.
By denoting
\[
\Phi_{m}(r)=-\frac{1}{|X_{m}|}\sum_{x\in X_{m}}\log\left(\frac{|\left\{ y\in X_{m}:\left\Vert x-y\right\Vert _{\infty}\leq r\right\} |}{|X_{m}|}\right),
\]
the \emph{approximate entropy} of the $N$ data points $u$ is
defined as
\[
\mathrm{ApEn}\left(m,r,N\right)=\Phi_{m+1}(r)-\Phi_{m}(r),
\]
with the convention that $\mathrm{ApEn}\left(0,r,N\right)=\Phi_{1}(r)$
and $X_{0}=\{\}$. The upshot of this is that small values of $\mathrm{ApEn}$
imply strong regularity (or persistence), whilst large values amount
to considerable irregularity in the time series $u$. From a concrete point of view,
approximate entropy is interpreted as a measure of randomness of a finite sequence. To complete
the geometric picture, observe that the \emph{Eckmann-Ruelle} entropy is indeed recovered in the {}small $r$, large $m$ limit:
\[
\underset{r\rightarrow0}{\lim}\,\underset{m\rightarrow\infty}{\lim}\,\underset{N\rightarrow\infty}{\lim}\mathrm{ApEn}\left(m,r,N\right).
\]
Amongst the practical uses of approximate entropy, for example when
studying time-series data of financial markets \cite{PK04} or heart
EEG data \cite{SD99,Joy00}, the literature often uses $m=1$ or $2$, together
with $r$ proportional to the standard deviation.\\

The combinatorial picture gives the key insight for estimating the
limiting distribution of approximate entropy and also identifying
sequences of extremal approximate entropy. Indeed, let $u\in\left\{ 0,1,...,S-1\right\}^{N}$
be a sequence of length $N$ on $S$ symbols (here we are taking $0<r<1$).
Let $\nu_{i_{1},...,i_{m}}$ be the frequency with which the block
$\left(i_{1},...,i_{m}\right)\subset\{0,1,...,S-1\}^{m}$ occurs in
$\tilde{u}=\left(u_{1},...,u_{N},u_{1},...,u_{m-1}\right)$.
This amounts to the frequency of the block in $u$ arranged around in a circle.
As before, set
\[
\widetilde{\Phi}_{m}=-\sum_{I\in\{0,1,...,S-1\}^{m}}\nu(I)\log\nu(I).
\]
The \emph{modified approximate entropy} is
\[
\widetilde{\mathrm{ApEn}}\left(m\right):=\widetilde{\Phi}_{m+1}-\widetilde{\Phi}_{m}
\]
so that its computation amounts purely to counting the relative frequencies
associated with every possible block of length $m$ occurring in the sequence.
This allowed Rukhin to get analytic proofs of the distribution of
$\mathrm{ApEn}$: for fixed embedding dimension $m$, we have then
\[
2N\left(\log S-\widetilde{\mathrm{ApEn}}\left(m\right)\right)\rightarrow\chi^{2}\left(S^{m+1}-S^{m}\right),
\]
so that the same behaviour follows for $\mathrm{ApEn}\left(m\right)$ because
\[
N\left(\mathrm{ApEn}\left(m\right)-\widetilde{\mathrm{ApEn}}\left(m\right)\right)=O_{\mathbb{P}}\left(\frac{1}{N}\right).
\]

\section{Approximate entropies of a network}
\label{sec_degree}

The \emph{degree sequence} of a (finite, unweighted, and undirected) network $G$ with
nodes labelled $\{1,...,N\}$ is $d=(d_{1},...,d_{N})$ where $d_{1}\geq...\geq d_{N}$
and $d_{i}=\deg(i)$ \cite{New10}. Note at this point that not all decreasing sequences of integers are
degree sequences of networks. The best known criterion was given by
Erd\H{o}s and Gallai \cite{TV03}: $d_{1}\geq...\geq d_{N}$ is the degree
sequence of a network with $N$ nodes if and only if $\sum d_{i}$ is even
and $\sum_{i=1}^{k}d_{i}\leq k\left(k-1\right)+\sum_{i=k+1}^{N}\min\left(d_{i},k\right)$
holds for every $1\leq k\leq N$. What this really says is that the trivial
bounds for the partial sums $d_{1}+...+d_{k}$ of the first $k$ largest
degrees, obtained by combining the contributions of $k(k-1)$ for the links
between those first $k$ and $\min\left(d_{k+1},k\right)+...+\min\left(d_{N},k\right)$
for the rest, is in fact optimal. In other words, whenever the condition is satisfied
for every $k$, there exists a network with such a degree sequence.
A constructive proof of this fact was given by Tripathi, Venugopalan
and West in \cite{TVW10}. In fact, Tripath and Vijay \cite{TV03}
observed that it is enough to check the condition for $k=n$ and every
$k$ such that $d_{k}>d_{k+1}$. \\

Good asymptotic bounds for the number of degree sequences of networks on $N$ nodes were given by Burns \cite{Bur07}; in this reference it was shown that the existence of constants $C_{0}$ and $C_{1}$ such that the number of degree sequences of graphs on $N$ vertices, $\Delta_{N}$, satisfies
$\frac{4^{N}}{C_{0}N}\leq\Delta_{N}\leq\frac{4^{N}}{\left(\log N\right)^{C_{1}}\sqrt{N}}$.
It is worth noting however that amongst the sensible candidates for
degree sequences, \emph{i.e.}, amongst monotone decreasing sequences of the
appropriate length, the proportion of these that are degree sequences
tends to zero, a fact conjectured by Wilf and proven by Pittel \cite{Pit99}.
\\

From now on, let $d_{1}\geq...\geq d_{N}$ be the degree sequence
of a finite network with distinct values $D_{1}>...>D_{s}$ such that
$D_{i}$ occurs $n_{i}$ times (so $N=n_{1}+...+n_{s}$). How best
to assign an approximate entropy to a degree sequence? We consider
three options in turn:
\begin{enumerate}
\item The simplest thing to do is just to compute the approximate entropy
of this monotone sequence viewed as a time series, due to a lack of
a natural ordering of nodes.
\item More sophisticated is to assign some combinatorial description of
the degree sequence, with potentially interesting entropic properties.
\item For networks with a natural ordering to their vertices, we can compute
approximate entropy of such an ordering. For example, a network that
has {}grown by a process of sequential node addition has a natural
ordering on the nodes according to when they were added. An example of this is the standard Barabási\textendash{}Albert model of preferential attachment \cite{BA99}.
\end{enumerate}

\subsection{The degree sequence as a monotone time series}

Option (1) does not seem to coincide with intuition of what an entropic
sequence should be: indeed if $m$ and $m+1$ divide $N$ then in
the small $r$ limit, the behaviour captures more the dimension than
the disorder in the sequence. This feature is highlighted by the following observation: suppose that $D_{i}-D_{i+1}>r$, then $\mathrm{ApEn}(m,r,N:m\mbox{ and }m+1\mbox{ divide }N)$
is maximal when $s=N/m$ and $n_{1}=...=n_{s}=m$ and minimal when
$n_{1}=...=n_{s}=m+1$. Let us sketch how to show this statement.
We may take $0<r<1$ and suppose the $D_{i}$ to be integers. Optimizing
$\mathrm{ApEn}(m,r,N)=\Phi_{m+1}-\Phi_{m}$ for a degree sequence
$d$ amounts to optimizing the ratio $R$ of geometric means of the
numbers of points in $X_{m}$ of $\left\Vert .\right\Vert _{\infty}$
distance $\leq r$ for each $x\in X_{m}$. Hence,

\[
R=\frac{\left(\prod_{x\in X_{m}}\left|\left\{ y\in X_{m}:\left\Vert x-y\right\Vert _{\infty}\leq r\right\} \right|\right)^{1/|X_{m}|}}{\left(\prod_{x\in X_{m+1}}\left|\left\{ y\in X_{m+1}:\left\Vert x-y\right\Vert _{\infty}\leq r\right\} \right|\right)^{1/|X_{m+1}|}}.
\]
In this case, $0<r<1$ implies that we need only optimize

\[
R=\frac{\left(\prod_{i=1}^{s}\Delta_{i}(m)^{\Delta_{i}(m)}\right)^{1/|X_{m}|}}{\left(\prod_{i=1}^{s}\Delta_{i}(m+1)^{\Delta_{i}(m+1)}\right)^{1/|X_{m+1}|}},
\]
\emph{i.e.},

\[
R=\prod_{i=1}^{s}\frac{\Delta_{i}(m)^{\Delta_{i}(m)/|X_{m}|}}{\Delta_{i}(m+1)^{\Delta_{i}(m+1)/|X_{m+1}|}}.
\]
Here $\Delta_{i}(m):=\max\left(1,n_{i}-\left(m-1\right)\right)$
such that $n_{1}+...+n_{s}=N$ and $1\leq s\leq N$. We are also free
to sort the $n_{i}$ such that $n_{1}\geq...\geq n_{s}$. Now, suppose
that there are $j\leq s$ separate $n_{i}$ such that $\Delta_{i}(m)=n_{i}-(m-1)>1$
and $k\leq j$ distinct $n_{i}$ such that $\Delta_{i}(m+1)=n_{i}-m>1$.
Set $r_{i}=n_{i}-m$ then $r_{1}\geq...\geq r_{j}\geq3>2=r_{j+1}=...=r_{k}>1\geq r_{k+1}\geq...r_{s}$.
The ratio $R$ now reads
\[
R=\prod_{i=1}^{j}\frac{r_{i}^{r_{i}/|X_{m}|}}{\left(r_{i}-1\right)^{\left(r_{i}-1\right)/|X_{m+1}|}}\prod_{i=j+1}^{k}4^{1/\left|X_{m}\right|}.
\]
The statement follows by writing out the behaviour of $x^{x/k}/\left(x-1\right)^{\left(x-1\right)/(k+1)}$
for $k$ an integer at least 2.

Thus this gives only an idea of proximity to such nonintuitive
entropic sequences. For example, with $0<r<1$ and $m=0$, the approximate
entropy is maximal for the sequence $\left(N-1,N-2,...,\left\lceil N/2\right\rceil,\left\lceil N/2\right\rceil,...,2,1\right)$; if $m=1$ and $N=9$, the maximum is realized by the sequence $\left(8,8,7,7,6,6,4,4,4\right)$.
A graph is regular if all its vertices have the same degree. The \emph{frequency partition} of a graph is a partition of its vertices grouped by their degrees. Such a notion is a graph invariant, but intuitively there are many non-isomorphic graph with the same frequency partition. It is known that every partition is a frequency partition of some graph, with the exception of $\left(1,1,...,1\right)$ (see \cite{rao}). A graph has a \emph{regular frequency partition} if each block of the partition is of the same size. In general, if $m=1$, the graphs that realize the maximum have degree sequence $(N-1,N-1,N-2,N-2,..., N-(N+1)/2,N-(N+1)/2,(N-1)/2)$ when $N$ is odd and $(N-1,N-1,N-2,N-2,...,N/2,N/2)$ when $N$ is even (we take $N \geq 4$).

\subsection{The slide of a degree sequence}

Let us now explore option (2). Given a degree sequence $d=\left(d_{1},...,d_{N}\right)$, for each
$i=1,...,N-1$ write down $0$ if $d_{i}=d_{i+1}$ and otherwise
write a string of $d_{i}-d_{i+1}$ $1$'s if $d_{i}>d_{i+1}$.
We denote this sequence by $\mbox{slide}(d)$. For a network $G$ with
degree sequence $d$, write $\mathrm{slide}(G)$. Note that for a
network $G$ on $N$ nodes, $N-1\leq\#\mathrm{slide}(G)\leq2(N-2)$
with the minimum length attained by (for example) regular networks
and the maximum attained by stars (\emph{i.e.} complete bipartite networks
with the singlet as a class of the bipartition).
Thus a degree sequence with $s$ distinct degrees is associated to
a binary sequence of $N-s$ zeros and $d_{1}-d_{N}$ ones and the
collection of associated sequences of networks with $N$ nodes is
a certain subset of binary sequences with up to $N$ zeros and up
to $N-1$ ones. For example, the degree sequence $(4,3,3,3,1)$ is
encoded as $\left(1,0,0,1,1\right)$. The associated binary code has
a simple interpretation in terms of {}``sliding'' down the degree
sequence: 0 means {}``go horizontally right'' and 1
means {}``continue going down''.
Not that not every binary sequence arises as the slide of some network. For
example, there is no network $G$ with $\mbox{slide}(G)=001$: indeed such a degree
sequence must be $d_{1}=d_{2}=d_{3}=d_{4}+1$ and none of the possibilities
$\left(3,3,3,2\right)$, $\left(2,2,2,1\right)$ or $\left(1,1,1,0\right)$
are graphical because they all have odd sum. Obviously the slide map
is not injective and further, networks of different numbers of nodes
can have the same slides.
The approximate entropy of binary sequences was studied by Pincus
\cite{Pin96} in the context of developing properties of normal
numbers. The notion of approximate entropy allows us to compare binary
sequences. The language used by Pincus to do so is as follows: a \emph{binary sequence} $u\in\{0,1\}^{N}$ of length $N$
is called \emph{$(m,N)$-random} if $\mathrm{ApEn}\left(m,N\right)(u)$
is maximal amongst all binary sequences of the same length. Let $m^{*}(N)$
be the largest integer such that $2^{2^{m^{*}(N)}}\leq N$, and call
a binary sequence $u$ \emph{$N$-random} if for $m=0,1,2,...,m^{*}(N)$
it is $(m,N)$-random. Finally, we can compare two binary sequences
$u$ and $v$ of the same length $N$ and say that $u$ is \emph{more
N-random} than $v$ if $\mathrm{ApEn}\left(m,N\right)(u)\geq\mathrm{ApEn}\left(m,N\right)(v)$
holds for all $m$ such that $1\leq m\leq m^{*}\left(N\right)$.

A characterization of binary sequences of the largest approximate
entropy was also given by Pincus: for $N\geq5$, the $N$-random binary sequences amount to
equivalence classes of sequences of length $N$ of a partially exchangeable
process in which {}``approximate stability of frequencies holds'',
in the sense that
\[
\left|\frac{\#\left\{ \left(a_{0},...,a_{m}\right)\mbox{-blocks in the sequence}\right\} }{N-m}-\frac{1}{2^{m+1}}\right|
\]
is as small as possible for each block type \textup{$\left(a_{0},...,a_{m}\right)\in\left\{ 0,1\right\} ^{m+1}$
and for every $0\leq m\leq m^{*}(N)$. }
That such sequences are asymptotically of large approximate entropy
can be seen immediately from Rukhin's characterization. What this
says is that the binary sequences of maximal approximate entropy amount
to optimal {}``truncations'' of normal numbers written in base 2.

If asymptotically all binary sequences arose as the
slides of some degree sequence, whilst the proportion of binary
sequences of length $n$ that are $n$-random tends to 0, we could expect
networks whose slides were arbitrarily nearby (it is unknown for which
$n$ there exist networks with $n$-random slides). In particular,
we conjecture the following result:
\begin{conjecture*}
The probability that a uniformly chosen slide of length $n$ is graphical
tends to 1 as $n\rightarrow\infty$.
\end{conjecture*}
A potentially fruitful approach to a proof include three steps: firstly, one may consider Pittel's approach for proving
the Wilf conjecture. The approach refines the insight of Erd\H{o}s-Richmond
in associating an integral estimate of the probability of surviving
the Nash-Williams graphicality condition. Subsequently, one should combine
the Kolmogorov 0-1 law with a sample family of networks whose slides
make an asymptotically non-zero contribution. An alternative is to
try show that with high probability, one can construct a network (perhaps
via the arguments of \cite{TVW10}) exploiting the {}``slack'' gifted
by the considerable non-uniqueness of slides. We have computed the
first few terms of the sequence $s_{2},s_{3},...,s_{6}$. The respective values are $3/4,3/8,13/16,20/32,58/64$.
We are now ready to define the \emph{slide entropy} of a network:
\begin{defn*}
If $G$ is a network with $N$ nodes then the \emph{slide entropy}
of $G$ is
\[
\mathrm{SlideApEn}(G):=\left(\mathrm{ApEn}(\mathrm{slide}(G),m,r)\right)_{m=1,...,m^{*}(N)}
\]
for any $0<r<1$.
\end{defn*}

This notion is the topic of the next section.

\section{Approximate entropy of slide sequences: application to static complex networks}

Scale-free networks are of interest due to their abundance in nature and technology \cite{BA99, New03, New10}.
Computationally, estimates of their behaviour as complex networks
are often made by treating their degree sequences as random variables
sampled from a scale free distribution $\pi\left(x\right)=\left(\gamma-1\right)x^{-\gamma}$
(for real $x>0$). A natural construction via growth through preferential
attachment was popularized by Barabási and Albert \cite{BA99}. The probability distributions this yields on networks of $N$
nodes (for each $N$) is quite distinct from simply sampling a scale
free distribution and trying to assemble a network from that, the so-called configuration model \cite{New03}. Scale
free networks are often compared with an older and very well studied
notion of random network, introduced by Erd\H{o}s and Rényi,
these are constructed by adding each edge with a fixed probability
$p$. For $N$ nodes, with $\lambda=Np$, their degree distribution
is asymptotically Poisson, $\pi(x)=\frac{\lambda^{k}e^{-\lambda}}{k!}$
($x>0$). Also of interest are random networks with exponential degree distributions, which naturally arise as the renormalization group fixed point of visibility graphs associated to random uncorrelated time series, with
$\pi\left(x\right)=\lambda e^{-\lambda x}$ $(x>0)$ \cite{Luq11, Luq09}.

\begin{figure}
\centering
\includegraphics[width=1\columnwidth]{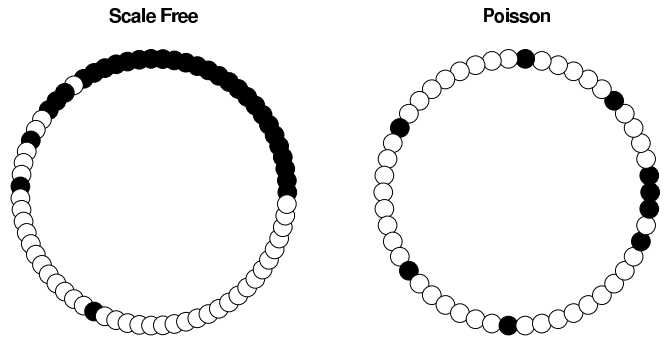}
\caption{Circularly arranged slides of random scale free
and Poisson networks of $N=50$ nodes illustrating typical structure.
Note that whilst $N=50$ is perhaps far too small to meaningfully
refer to the distribution as scale free the figure is intended
only to be illustrative of the general appearance of the slides.}
\label{circles}
\end{figure}

The approximate entropy of a generic slide with $d_{1}-d_{N}$
ones and $N-s$ zeros can be computed by considering the Markov chain
with state space $X=\left\{ 0,1\right\} $ and transition matrix $P$
given by $P_{00}=P_{10}=p$ and $P_{11}=P_{01}=q=1-p$ (thus $\pi(0)=p$
and $\pi(1)=q$). We recall the following result of Pincus.
For a first order stationary Markov chain with discrete state
space $X\subset\mathbb{N}$, transition probabilities $P_{xy}=\mathbb{P}(\mbox{travel from }x\mbox{ to }y)$
and stationary distribution $\pi$ with $r<\min_{x\neq y\in X}\left|x-y\right|$
then almost surely for every $m$
\[
\underset{N\rightarrow\infty}{\lim}\mathrm{ApEn}\left(m,r,N\right)=-\sum_{x,y\in X}\pi\left(x\right)P_{xy}\log\left(P_{xy}\right)
\]
It follows that
\[
\mathrm{SlideApEn}=H\left(\frac{N-s}{N-s+d_{1}-d_{N}}\right),
\]
where as usual $H\left(p\right)=-\left(p\log p+\left(1-p\right)\log\left(1-p\right)\right)$.
However for a scale free network on $N$ nodes with degree distribution sampled from a truncated scale free distribution
$\tilde{\pi}(k)\sim k^{-\gamma}$ for $k\in\{1,...,N-1\}$. In the
large $\gamma$ limit the network is dominated by degree 1 nodes as
we can expect to have at least $Nk^{-\gamma}/\zeta(\gamma)$ nodes
of degree $k$ which tends to zero for all $k$ except $k=1$. This
creates large regions of zeros in their slides. Assuming generic
behaviour in the region of the slide not accounting for the degree
$1$ nodes, the $\widetilde{\mathrm{ApEn}}$ estimate of Rukhin suggests
intuitively that the slide entropy should be roughly monotone decreasing
in $\gamma$ for each fixed $N$. This intuition can be seen numerically
in Fig. IV.1.
Estimating $d_{1}-d_{N}$ can be done by using the following elementary
result. Let $X_{1},...,X_{N}$ be drawn from a probability distribution
$\pi$ on $\mathbb{R}$ with cumulative density $\Pi$. Then $E_{m}:=\mathbb{E}\left(m^{\mathrm{th}}\mbox{ largest }X_{i}\right)$
is given by
\[
\int_{\mathbb{R}}x\pi(x)\frac{N!}{\left(N-m\right)!\left(m-1\right)!}\Pi(x)^{m-1}\left(1-\Pi(x)\right)^{N-m}dx.
\]
If we are drawing the degrees $\deg(v)$ of nodes $v\in\left\{ 1,...,N\right\} $
then $\mathbb{E}(d_{1}-d_{N})=E_{1}-E_{N}$ can be computed for the
scale free degree-distribution (we have $E_{N}=1$ almost surely) given approximately
(following Ghoshal-Barabási \cite{GB11}) by
\[
E_{1}\approx\left(N-1\right)^{1/\left(\gamma-1\right)}\Gamma\left(\frac{\gamma-2}{\gamma-1}\right)
\]
This becomes a good approximation for large $N$, but performs poorly
near $\gamma=2$ due to the pole at 0. To estimate the expected number
of distinct degrees $s$, we associate to the continuous probability
distribution $\pi$ on $\mathbb{R}$ a distribution $\overline{\pi}$
on $\mathbb{N}$. An elementary argument says that the expected number
of distinct values of a random sample of size $N$ from the distribution
$\overline{\pi}$ is
\[
s(N)=\sum_{n\in\mathbb{N}}\left(1-\left[1-\overline{\pi}(n)\right]^{^{N}}\right).
\]
In this way, for a sample of $N$ random variables from a scale free distribution
$\overline{\pi}(n):=n^{-\gamma}/\zeta\left(n\right)$ and upon comparing
with the integral, we get the following estimate
\[
s(N)\approx{}_{2}F_{1}\left(-\frac{1}{\gamma},-N;\frac{\gamma-1}{\gamma};\frac{1}{\zeta(\gamma)}\right).
\]
This provides an analytic expression for the entropy of a generic
slide of the same 0-1 distribution, but is a considerable overestimate
for real scale free slides. The slight {}``kink'' in Fig IV.2. around
$\gamma\approx2.4$ arises from the change in generic behaviour.
We similarly obtain generic estimates for exponential degree sequences
by computing $E_{1}-E_{N}=\frac{1}{\lambda}\sum_{i=1}^{N-1}\frac{1}{i}\sim\log\left(N-1\right)+O\left(\frac{1}{N}\right)$
and $s(N)\approx\int_{1}^{\infty}\left(1-\left(1-\lambda e^{-\lambda x}\right)^{N}\right)dx$.
For Poisson networks, we find numerically that the generic estimate
provides a good approximation.
It is interesting to ask which probability distributions $\overline{\pi}$
on $\mathbb{N}$ tend to give rise to networks of the largest and
smallest slide entropies. Amongst such distributions on $N$ nodes,
we expect that the uniform distribution on $\{0,1,...,N-1\}$ is of
the greatest typical slide entropy. Note that the point distribution
on any fixed $k\in\{0,1,...,N-1\}$, such that the networks desired
exist, always gives rise to $0$ slide entropy networks.

\begin{figure}
\begin{center}
\includegraphics[width=0.8\columnwidth]{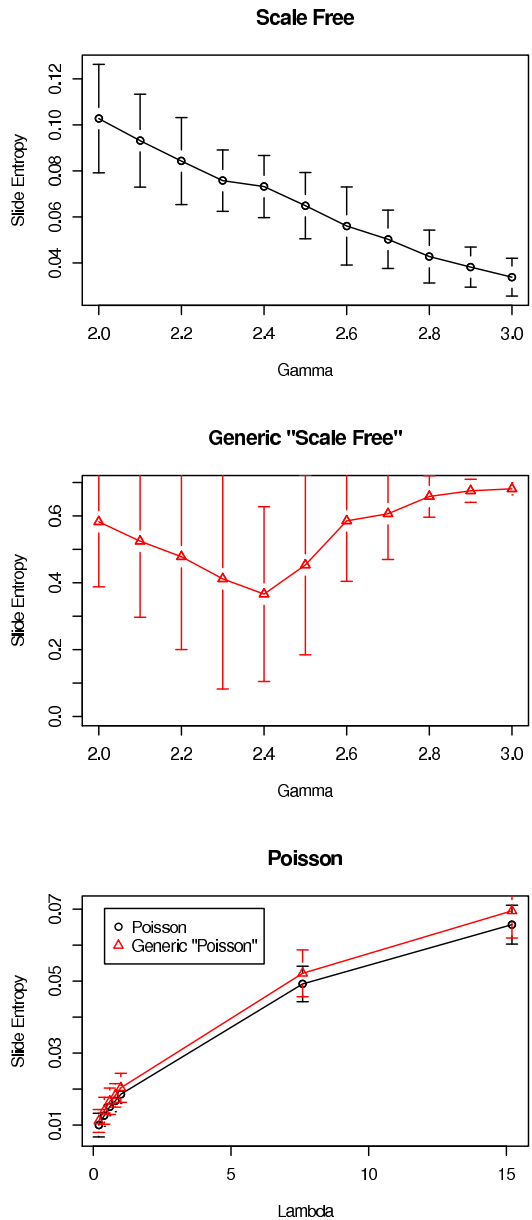}
\caption{Slide entropies for scale free and Poisson (Erd\H{o}s\textendash{}Rényi)
networks of $N=2000$ nodes: generic values given are those
of uniform random slide entropies of slides with the same distribution
of 0s and 1s. Error bars indicate two observed standard deviations.}
\label{}
\end{center}
\end{figure}

\section{Approximate entropy of growing networks: horizontal visibility graphs}
\begin{figure*}
\begin{center}
\includegraphics[width=0.75\textwidth]{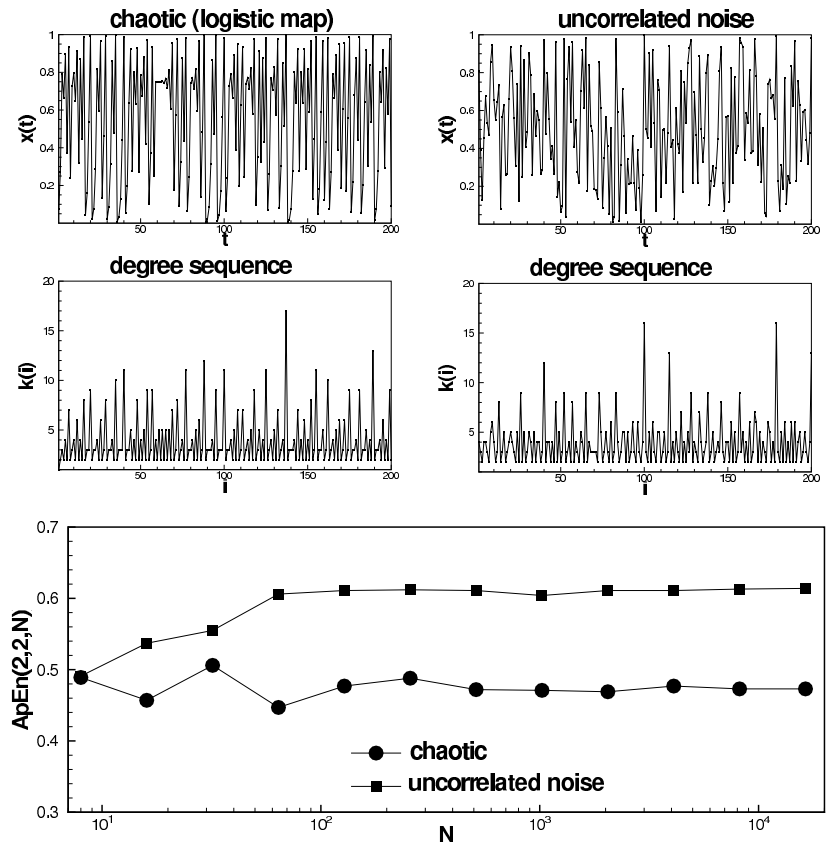}
\caption{\emph{Top}: Sample series extracted from (left) the fully chaotic logistic map $x_{t+1}=4x_t(1-x_t)$ and (right) uncorrelated random variables from $U[0,1]$. Below, we show a sample of the degree sequences of their associated HVgs. \emph{Bottom}: Values of ApEn$(2,2,N)$ of each  HVg, as a function of the series size $N$. Note that the time is in logarithmic scales. The values for
the graph associated to the noisy process are larger than those associated to the chaotic process, in concordance with the entropy associated to the underlying dynamical process.}
\label{visib}
\end{center}
\end{figure*}

Within nonlinear time series analysis, the so called \emph{visibility algorithms} \cite{Lac08,Lac09,Lac10,Luq09,Luq11}
are a family of methods that directly map a given time series of $N$ data into a network of $N$ vertices (a so called visibility graph),
where the edge set is constructed according to specific geometric criteria to be applied among the data set.
In previous works it has been shown that the associated visibility graph of a time series with a given information is conserved or inherited
in the topology of the associated visibility graph, including nontrivial structures such as chaotic or fractal dynamics.
To cite a few, within the so called visibility algorithm approach \cite{Lac08,Lac09}, series extracted from a fractional Brownian motion with Hurst exponent $H$ map into scale-free visibility graphs with degree distribution $P(k)\sim k^{-\gamma}$, where the linear relation $\gamma=3-2H$ quantitatively relates the structure of the dynamical process ($H$) with the topology of the associated graph ($\gamma$). Within an alternative approach coined as the horizontal visibility algorithm \cite{Luq09,Gut11}, it was shown \cite{Lac10} that horizontal visibility graphs distinguish between correlated stochastic,
uncorrelated and chaotic processes, and in each of these cases the visibility graph has exponential degree distribution $P(k)\sim e^{-\lambda k}$ with the value of $\lambda$ characterizing the particular process. Recently, it has been also suggested that the Shannon entropy over the degree distribution of a horizontal visibility graph is a first order approximation to the Kolmogorov-Sinai entropy of the associated dynamical system \cite{Luq11}.\\
As a first comment, note that the visibility graph associated to a given time series conserves, by construction, the temporal ordering of the data, \emph{i.e.} temporal correlations amongst the data. This is due to the fact that in the mapping algorithm, each datum $x_i$ maps into a labeled vertex $n_i$, that is to say, a natural ordering of the vertex set emerges, respecting the temporal correlations in the series. The implications are twofold: (i) within (horizontal) visibility graphs one has a natural ordering of the degree sequence, which allows us to unambiguously calculate the approximate entropy of such series, and (ii) since that ordering is related to the temporal correlations of the associated series, the approximate entropy of a visibility graph may provide a measure of the associated series complexity, that is, it becomes a measure directly related to the original one introduced by Pincus in the framework of dynamical systems. In other words, whereas our previously defined, network-based, slide entropy accounts for the heterogeneity of the network itself (with no dynamical/temporal information whatsoever), in the context of visibility graphs this network-based measure is indeed capturing some dynamical information. Notice also that since each datum in a time series is associated to a labeled vertex in the graph, one can view a visibility graph as a dynamically growing network: as time evolves, the dynamical process generates a trajectory (time series) whose associated visibility graph grows. The approximate entropy of its degree sequence accounts for the information stored in the network growing process.\\
In order to test the aforementioned conjectures, we will address, within the so called horizontal visibility algorithm, three types of time series whose associated approximate entropy, associated to the amount of information needed to unravel the underlying dynamics, is qualitatively different: periodic series (\emph{i.e.} regular dynamics with point-like attractor measure), chaotic series (deterministic dynamics with finite attractor measure) and
white noise (stochastic dynamics with infinite attractor measure). We proceed as follows: let $\{x_t\}_{t=1,. . .,N}$ be a real-valued time series of $N$ data. The horizontal visibility algorithm assigns each datum of the series to a vertex in
the {\sl horizontal visibility graph} (HVg). Then, two vertices $i$ and
$j$ in the graph are connected if one can draw a \emph{horizontal}
line in the time series joining $x_i$ and $x_j$ that does not
intersect any intermediate data height. Hence, $i$ and $j$ are two connected nodes if the following
geometrical criterion is fulfilled within the time series:
\begin{equation}
x_i,x_j > x_n, \ \forall \ n \ \left | \ i < n < j\right.  \label{criterio}
\end{equation}
The generated HVg has a degree sequence of the kind $\{k_1,k_2,...,k_N\}$, where $k_i$ is the degree of vertex $i$, that is to say, associated
to datum $x_i$ in the original series (as opposed the definition in section \ref{sec_degree}, not that this degree sequence is not monotonically decreasing since it has a natural ordering already explained above). We finally calculate our network based ApEn. Results for periodic, chaotic and noisy series are shown, for specific values of the ApEn parameters, are summarized in table \ref{table1} and in Fig. \ref{visib}. On this respect, we can highlight the following comments:
\begin{table}
\begin{ruledtabular}
\begin{tabular}{cccccddd}
\textbf{Series description}&ApEn$(2,2,2^{14})$\\
\hline
periodic series (T=2)&0.001\\
chaotic series (logistic map, $\mu=4.$)&0.47\\
$U[0,1]$ uncorrelated&0.62\\
\end{tabular}
\end{ruledtabular}
\caption{\label{table1} Values of ApEn for concrete parameters $m=2$, $r=2$ and size $N=2^{14}$, of the HVg associated to three types of time series: (i) a periodic series of period 2 (deterministic dynamics with an underlying attractor of zero measure), (ii) a chaotic series extracted from the fully chaotic logistic map $x_{t+1}=4x_t(1-x_t)$ (deterministic dynamics with an underlying attractor of finite measure), and (iii) a series of uncorrelated random variables extracted from a uniform distribution $U[0,1]$ (stochastic dynamics, \emph{i.e.} dynamics with an hypothetical infinite-dimensional attractor). The approximate entropy of the visibility graphs increase as a function of the associated series information.}
\end{table}

\begin{itemize}
\item{HVg associated to periodic series}: by construction,
these networks have a very homogeneous structure \cite{Luq09}, which can indeed
by seen as a concatenation of a network motif, the structure of this
network motif being intimately related to the series periodicity.
Accordingly, their ApEn is small, having a vanishing value for large
embedding dimension $m$. If we make use of small values of $m$,
the ApEn statistic can be used to distinguish several degrees of periodicity,
associated to the the heterogeneity of the visibility network root
motif.
\item{Visibility graphs associated to random uncorrelated white noise}:
white noise is a maximally entropic signal according to any well defined
information theoretic measure. In a previous work \cite{Luq11}, it was shown that
the Shannon entropy over the degree distribution of an HVg is indeed maximized for uncorrelated white noise.
Here, unlike periodic series, results are
close to convergence for $m=2$. For a given series size $N$, noise
yields the maximal ApEn, but this value increases with the series
size $N$, as it should.
\item{Visibility graphs associated to chaotic maps}: The ApEn of the
associated networks reach a non zero value, reminiscent of the underlying
attractor of the dynamics. Convergence is reached for $m=2$ and,
unlike noise, convergence as a function of series size is also reached
here, as it should.
\end{itemize}
This preliminary analysis suggests that the network structure captures the
inherent complexity of the associated time series. This supports the
well-posedness of the visibility graphs as a tool for time series
analysis. Based on this conclusion, we can also point out that the
network theoretical variation of the approximate entropy statistic
can effectively distinguish different network structures according
to their associated 'complexity'.\\

\emph{Mixed statistic:} So far, two alternative ways have been considered in order to define the sequence
over which the approximate entropy is computed, namely (i) networks without a predefined time arrow:
computation is performed over the slide of the (monotonically decreasing) degree sequence, and (ii) growing networks: computation is performed over
the (time ordered) degree sequence. A mixed approach
consists in computing ApEn over the slide of a time ordered degree sequence: we consider an
example of this next.

\section{Approximate entropy and biological networks: an application to cancer genomics}

Finally, we focus on a potential application of these ideas to the field of cancer genomics. A key feature of cancer genomes is the abnormal copy number of genes. Since healthy cells are diploid they have 2 copies of each gene, however, in cancer cells, genes are deleted or may be present in multiple copies. Genes also have a natural ordering since they can be located to specific positions on the genome. Thus, for each tumour we can measure a copy-number profile along the genome. For technical reasons this is represented as a continuous valued variable (segmented data) \cite{Chi07}, with neighboring genes more likely to have the same value. This copy-number profile varies along the genome, and can therefore be mapped as a time series, where genomic position plays the role of time. Thus, an HVg can be constructed for this genomic series of copy number values. We hypothesized that this HVG construction could encapsulate important information concerning the distribution and shape of the copy-number profiles of each individual tumour, a hypothesis that we test a posteriori by correlating the resulting entropy scores to known cancer phenotypes.

As a data set we considered the copy-number data of 171 breast cancer patients \cite{Chi07}, for which three phenotypic categories were available: estrogen receptor status (ER), whether the patient's tumour metastasized or not (DM) and histological grade (3 levels represents levels of differentiation from normal healthy tissue). For each tumour we computed the slide and ordered entropies from the HVg graphs and asked if these differed between phenotypes.

For brevity, we shall say that two phenotypes are \emph{distinguished} by some associated quantity if the means of the quantity for each phenotype are statistically significantly different (say at the 5\% level). Interestingly, we find increases in both entropies as the grade of the cancer increases and in the case of distal metastasis (DM). Moreover, the slide entropy distinguishes grade 1 breast cancer from grades 2 and 3 (Welch $t$-test with $p$-value 0.026 between grade 1 and grade 2 and $p$-value 0.006 between 1 and 3), however the ordered approximate entropy of the degree sequence of the HVg distinguishes grades 1 and 2 from 3 ($p$ value < 0.001 between grade 2 and 3). Both slide and HVG-ordered entropies distinguish
ER negative (0) and ER positive (1) breast cancer whilst neither distinguished DM-0 from DM-1. These results show that the ApEn and HVg construction can indeed capture interesting clinico-pathological of cancer genomes.

\begin{figure*}
\begin{center}
\includegraphics[width=2.0\columnwidth]{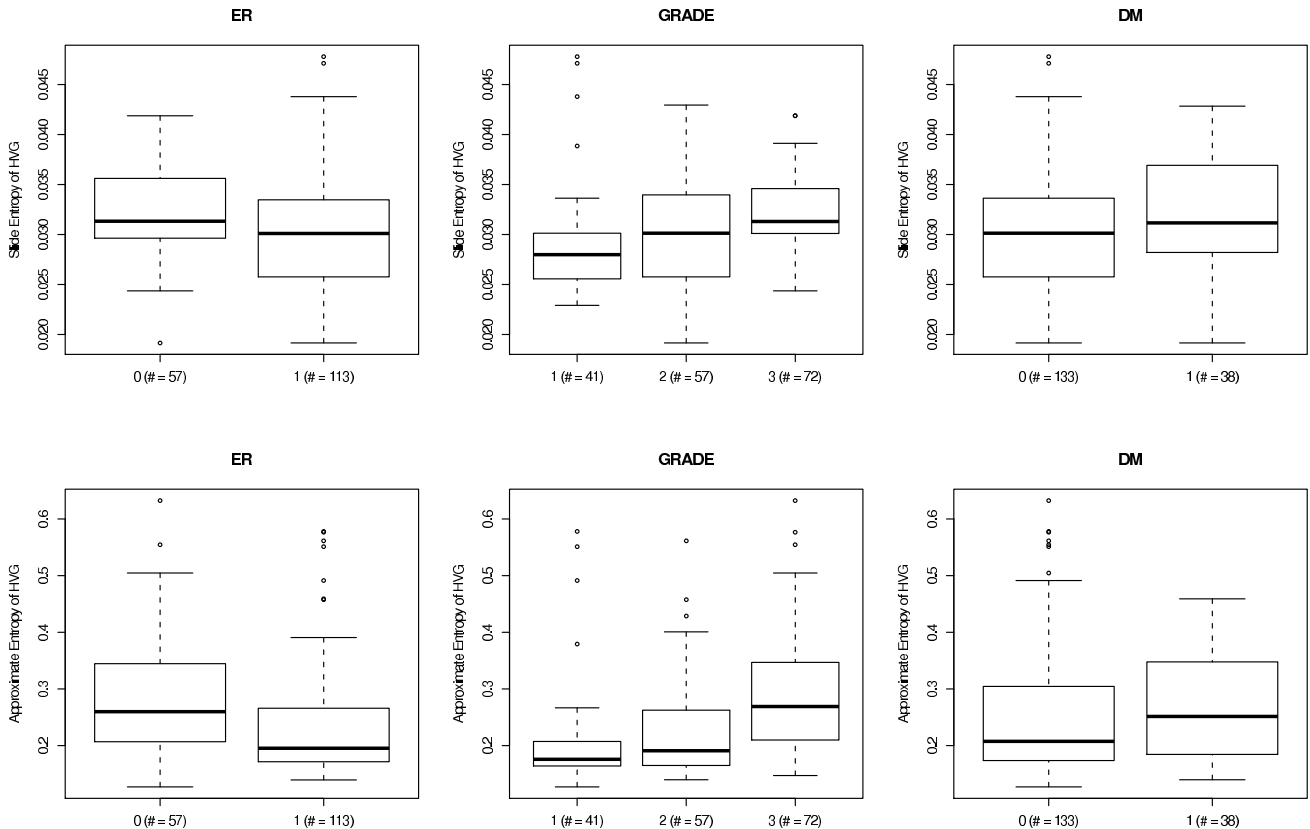}
\caption{Slide entropy ($m=1$) and ordered degree sequence
entropy ($m=1$ and $r<1$) of degree sequences of HVGs constructed
from copy number data of 171 breast cancer patients split according
to phenotype. ER denotes estrogen receptor status, indicating whether
the cancer cells depend on estrogen for their growth, GRADE indicates
the extent to which the cells have differentiated away from being
normal cells and DM denotes whether or not the patient has suffered
distal metastasis (\emph{i.e.} whether the cancer has spread).
The bars represent $5$ and $95\%$ quantiles of the distribution.}
\label{cancer}
\end{center}
\end{figure*}

We conclude this section by mentioning that the ordinary approximate entropy of the
copy number data itself (segmented and viewed as a time series) can also distinguish
grade 1 and 2 from grade 3 breast cancer but not grade 1 from grade 2. Our analysis is reported in Fig. \ref{cancer}.

\section{Conclusions}
The original definition of approximate entropy quantifies the structure of a system's underlying phase space by looking at time evolving trajectories
over such phase space, hence it requires a time ordered sequence. If the system under study is a network, its structure can still be studied with a use of approximate entropy, but it obviously requires some modifications/assumptions. If the network under study is generated by a dynamical process, then a natural time ordering can be defined over the vertex set, and ApEn is naturally extended to this domain: such is the case of visibility graphs, Barabasi-Albert or in general any kind of dynamically growing networks. Conversely, if the network under study is static, the natural extension of ApEn is not straightforward. In this case, we can still compute approximate entropy of various networks parameters, but this requires to make some choices. In this work we have computed the approximate entropy of parameters related to the degree sequence. In the attempt of capturing only the relevant information, we have introduced the slide sequence of a network. We have shown that approximate entropy permits to distinguish between the usual ensembles of Poisson and scale-free networks.

\begin{figure}
\begin{center}
\includegraphics[width=0.8\columnwidth]{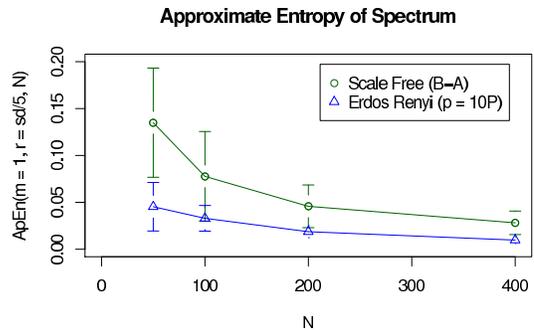}
\caption{The approximate entropy of the adjacency spectrum: the figure illustrates ApEn for the Barabasi-Albert scale free networks ($\gamma = 3$) and for ordinary Erd\H{o}s\textendash{}Rényi graphs, where the probability of each edge is 10 times the critical probability for the graph to be (almost surely) connected.}
\label{spectrum}
\end{center}
\end{figure}

Moving from static to dynamical networks, we have focused on horizontal visibility graphs. Our findings suggest that the approximate entropy of these objects is intimately related to the amount of information that the underlying dynamical process is generating. Indeed, we have given further evidence that just the degree sequence of an HVg have the power of discriminating between different dynamical processes. Finally, we have applied our statistic to specific HVg generated by genomic data. We have considered cancer networks because their amount of disorder measured by entropic quantities has already been utilized as a classifier \cite{TS11}.  The apparent capacity of this statistic to distinguish between several degrees of cancer should be certainly clarified in further work. It is a challenge to observe analogues of phase transitions in statistics related to this type of data. Let us briefly conclude with two open problems. The first problem is about giving an interpretation to the approximate entropy as defined here; the second one is about identifying naturally ordered sequences associated to networks.
\begin{itemize}
\item What does the slide ApeN tell us about a network? If we take a series and measure the ApEn of its visibility graph, what information do we learn from the dynamical process that generated the series? There should be a neat correlation between the ApEn of the series and the one of the visibility graph. Such a correlation may be used to determine what kind of information about a series is not seen by the visibility graphs approach. The gain is to uncover the limits of methods for time series analysis based on visibility graphs.
\item We have already mentioned that ApEn makes sense only when applied to an ordered list of numbers obtained from a network. The first intuitive choice was the degree sequence when arranged in the nonincreasing order. Clearly, this is not only the most natural choice, but it is also the easiest one. There are many potential generalizations based on different criteria. From a dynamical viewpoint, one could label the vertex set and then generate random walks over a given network. These are analogous to run trajectories over a dynamical system's phase space. ApEn is then computed over these trajectories.

    From a combinatorial viewpoint, the most straightforward generalization consists of looking at the second, third neighbours, and so on. These are sometimes called the \emph{shells} of a node. In this way, we may consider ApEn of the sequences generated by the number of (deterministic) walks of a given length. A variety of choices is then available: to look at the sequence of the number of walks of growing length, to count the number of walks starting from different nodes, \emph{etc.}. Remarkably, this allows us to associate various sequences to each node, which can be then compared, averaged, \emph{etc.}. The gain is the possibility of introducing network parameters for quantifying the \emph{disorder} in the cycle structure of the graph. Graphs with a particularly disordered cycle structures, like for example the controllable graphs introduced in , are expected to have higher ApEn of their walks.

    If instead of combinatorial criteria, we aim at a more algebraic perspective, a first choice consists of taking the spectrum of a matrix that represents the network faithfully, like the adjacency matrix or a Laplacian. Indeed the spectrum of a network is a graph invariant and a naturally ordered sequence. A very superficial analysis based on Fig.\ref{spectrum} suggests that ApEn does not contain valuable information, or at least information that is not easy to interpret. Hence, it remains an open problem to determine what kind of network properties are identified by computing ApEn of spectra and if this quantity grasps something about different matrix ensembles.
    \end{itemize}

\section*{Acknowledgements}
JW is supported by EPSRC through a CoMPLEX PhD studentship. LL acknowledges financial support from the MEC and Comunidad de
Madrid (Spain) through projects FIS2009-13690 and S2009ESP-1691. SS is
supported by the Royal Society and AET by a Heller Research Fellowship.

\end{document}